# Ultra-broad near-infrared photoluminescence from crystalline (K-crypt)$_2$Bi$_2$ containing [Bi$_2$]$^{2-}$ dimers

Hong-Tao Sun,*[a] Tetsu Yonezawa,[a] Miriam M. Gillett-Kunnath,[b] Yoshio Sakka,[c] Naoto Shirahata,[c,d] Sa Chu Rong Gui,[e] Minoru Fujii,[e] and Slavi C. Sevov*[b]

**For the first time, we report that a single crystal of (K-crypt)$_2$Bi$_2$ containing [Bi$_2$]$^{2-}$ displays ultra-broad near-infrared photoluminescence (PL) peaking at around 1190 nm and with a full width at the half maximum of 212 nm, stemming from the inherent electronic transitions of [Bi$_2$]$^{2-}$. The results not only add to the number of charged bismuth-based species with luminescence, but also deepen the understanding of Bi-related NIR emission behavior and lead to the reconsideration of the fundamentally important issue on Bi-related PL mechanisms in some material systems such as bulk glasses, fibers, and conventional optical crystals.**

Heavier p-block elements could form fascinating structures such as anionic and cationic polyhedra.[1-4] Bismuth, which has been called 'the wonder metal', is one of the most thoroughly investigated member of these elements.[1-5] So far, a broad range of compounds containing bismuth polycations or polyanions have been successfully synthesized.[1-4] In contrast to this rapid advance in the synthesis of novel bismuth structures, their optical properties, especially their photoluminescence (PL), have not attracted the attention they deserve. Recently, Sun *et al.* found that molecular crystals containing bismuth polycations such as Bi$_5^{3+}$ and Bi$_8^{2+}$ exhibit extremely broad near-infrared (NIR) photoluminescence,[4a-4c] suggesting that such systems with bismuth polycations have great potential for photonic applications.[4] Interestingly, very recently it was found that substructures of Bi$^+$ stabilized by zeolite Y frameworks also demonstrate peculiar NIR emission bands because of their intrinsic electronic transitions.[5] Insights obtained from the determination of the Bi distribution in these porous structures by high-resolution synchrotron powder X-ray diffraction coupled with detailed PL evaluation and quantum chemistry calculation of the Bi$^+$ substructures have inspired great interest in the exploration of more generic evolution rules of the Bi oxidation states. This has raised new possibilities for the design and synthesis of novel photonic materials using charged elemental clusters as the optically active centers.[5a] However, all aforementioned Bi species were positively charged.[4,5] As far as we are aware, up to now the PL properties of Bi polyanions have not been demonstrated experimentally, although a number of crystalline compounds containing such peculiar structural units have already been reported.[3] It is noteworthy that many of these compounds with Bi polyanions absorb light in the visible range of the electromagnetic spectrum since they are colored, thus giving promise for unique optical properties. It is reasonable, therefore, to anticipate that some of the polyanions may turn out to be novel optical emitters, perhaps in important spectral ranges such as biological and/or telecommunication optical windows, due to their inherent electronic transitions.

In this communication, for the first time we report that a single crystal of (K-crypt)$_2$Bi$_2$ which contains [Bi$_2$]$^{2-}$ polyanions displays ultra broad NIR emission at around 1190 nm with a full width at the half maximum (FWHM) of 212 nm. The single crystal was characterized by single-crystal X-ray diffraction (XRD), diffuse reflectance spectroscopy, and PL measurements. Our results revealed that [Bi$_2$]$^{2-}$ is a NIR emitter resulting from the characteristic electronic transitions from the excited levels to the ground level. The experimental results reported here require the reconsideration of the fundamentally important issue on Bi-related PL mechanisms in some material systems such as bulk glasses, fibers, and conventional optical crystals.

*[a] Division of Materials Science and Engineering, Faculty of Engineering, Hokkaido University, Kita 13, Nishi 8, Kita-ku, Sapporo 060-8628, Japan. Fax: +81-11-706-7881; E-mail: timothyhsun@gmail.com*
*[b] Department of Chemistry and Biochemistry, University of Notre Dame, Notre Dame, Indiana 46556, USA. Fax: +1 (574) 631-6652; E-mail: ssevov@nd.edu.*
*[c] Advanced Ceramics Group, Materials Processing Unit, National Institute for Materials Science (NIMS), 1-2-1 Sengen, Tsukuba-city, Ibaraki 305-0047, Japan*
*[d] PRESTO, Japan Science and Technology Agency (JST), 4-1-8 Honcho Kawaguchi, Saitama 332-0012, Japan*
*[e] Department of Electrical and Electronic Engineering, Kobe University, Kobe 657-8501, Japan*

(K-crypt)$_2$Bi$_2$ crystal was first synthesized and structurally characterized by Xu *et al.* using the precursor of K$_3$Bi$_2$.[3a] Here, we used an alternate approach to obtain high-quality single crystals (see details in Supporting Information). In brief, an ethylenediamine solution of 2,2,2-crypt is pipetted onto powdered precursor of K$_5$Bi$_4$ and then stirred for 15 minutes while forming a bright green-blue solution.[6] After filtering, the filtrate is carefully layered with three parts toluene followed by two parts hexane. After several days, dark red-brown plates, cubes, and blocks of (K-crypt)$_2$Bi$_2$ cleanly crystallize. Representative crystals of the different morphologies were indexed by single-crystal X-ray diffraction, and all of them display the same unit cell as the that of the previously reported (K-crypt)$_2$Bi$_2$, thus confirming the structure of exactly the same compound.[3a] The obtained high-quality single crystals without impurities paves the way for the following spectroscopy evaluation. The naked [Bi$_2$]$^{2-}$ dianion is oriented along the 3-fold axis of the rhobohedral structure and its charge is balanced by two [K-crypt]$^+$ cations (Figure 1). Due to the air- and moisture-sensitivity of the compound, its UV-vis-NIR absorption spectrum was measured on crystals enclosed between two 1 mm thick pure silica pieces (V-570 spectroscope equipped with an integrating sphere, JASCO, Japan). Steady-state PL measurements were carried out at room temperature with 641 nm excitation light from a laser diode. The signal was analyzed by a single grating monochromator and detected by a liquid-nitrogen-cooled InGaAs detector.

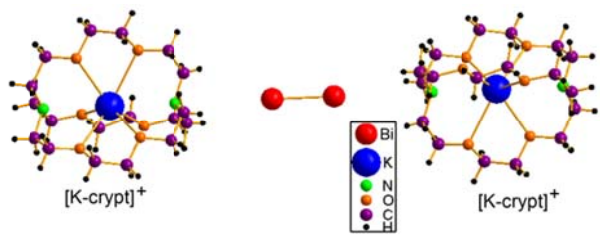

**Figure 1.** Part of the structure of (K-crypt)$_2$Bi$_2$.

The UV-vis-NIR diffuse reflectance spectrum demonstrates peculiar absorption bands at 471, 593, 705, 990, 1090, and 1273 nm (Figure 2). It has been shown before that [Bi$_2$]$^{2-}$ in an ethylenediamine solution shows two broad peaks at 435 and 605 nm without observable NIR bands.[3c] One possible rationale for this difference might be that the bond length of [Bi$_2$]$^{2-}$ in the solid state is somewhat different from that in solution since the polarization of the valence electron density distribution in the dianion by the surrounding cations in the solid state will effectively reduce the intra-molecular electrostatic repulsion and hence shorten the Bi-Bi bond length.[3c] It is also noted that the absorption in the NIR region is much weaker than that in the visible range, suggesting that the electronic transition probabilities in the NIR and visible ranges are different.

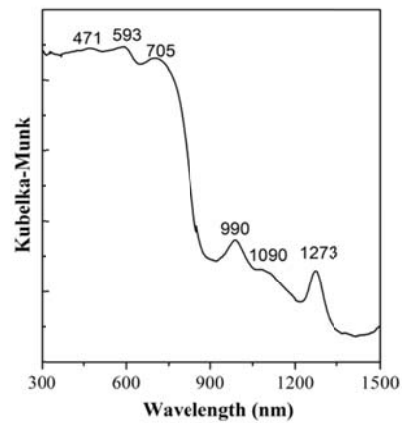

**Figure 2.** The UV-vis-NIR diffuse reflectance spectrum of crystalline (K-crypt)$_2$Bi$_2$.

Next, we studied the PL properties of the crystalline (K-crypt)$_2$Bi$_2$. As shown in Figure 3, the emission spectrum generated with 641 nm excitation light has a very broad emission range, spanning from 975 to 1400 nm. The FWHM of the spectrum is 212 nm and the emissoin peak is at around 1190 nm. Furthermore, it is found that the spectrum does not show the Gaussian profile, but rather two notable shoulders at *ca.* 1050 and 1330 nm are present as well. We further analyzed the emission spectrum by Gaussian decomposition, resulting in three decomposed curves peaking at 1047, 1190 and 1331 nm with FWHMs of 99, 146, and 56 nm, respectively. It is necessary to point out that these three emission energies are lower than the corresponding absorption energies at 990, 1090, and 1273 nm, respectively (Figure 2). The difference between positions of the band maxima of the absorption and emission spectra of the same electronic transition is due to the loss of excitation energy by thermalization of the excess vibrational energy. In general, there are two possibilities for the emissions under 641 nm excitation: they are either single- or multiple-photon processes. In the latter case, the active center will sequentially absorb at least two photons to emit one photon. In order to ascertain which process contributes to the observed emission, we ploted the log–log dependence of the NIR emission intensity at peak wavelength as a function of the excitation power. As displayed in Figure 4, the slope is unity, thus evidencing that the emission is a single-photon process. It is noteworthy that the sample shows similar emission behaviors under the excitation of 514.5 nm (Supporting Information, Figure S1), further indicating that the PL stems from the same type of active center.

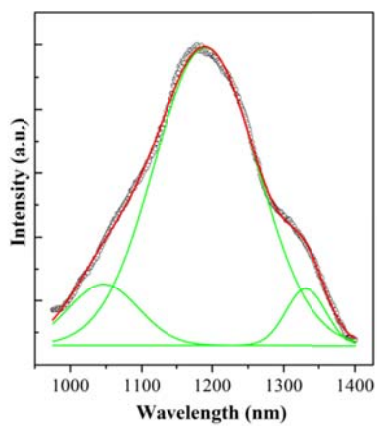

**Figure 3.** The PL spectrum of (K-crypt)$_2$Bi$_2$ crystals under 641 nm excitation light. The black, red and green curves are experimental, fitted, and three decomposed Gaussian peaks, respectively.

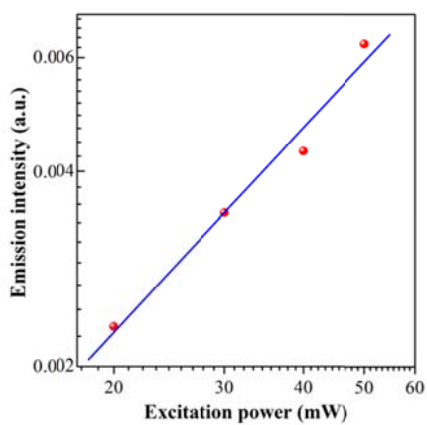

**Figure 4.** Log–log dependence of the emission intensity at the peak wavelength on the excitation power of the 641 nm diode laser.

It is well known that relativistic effects could give rise to a very strong spin-orbit coupling and exert a profound influence on the chemical bonding of compounds involving bismuth.[3c,5a,7,8] This makes it rather difficult to accurately determine the excitation energies of bismuth related species. For instance, Dai *et al.* first calculated the theoretical excitation energies of [Bi$_2$]$^{2-}$ with $d$(Bi-Bi) = 2.8377 Å in (K-crypt)$_2$Bi$_2$ using the Beijing density function program, and revealed that the anion should have at least four excitation bands located at around 1380, 1062, 562, and 399 nm.[3c] In a more recent work, Sokolov *et al.* calculated the excitation energies using the Gamess (US) quantum-chemical code.[8] The authors showed that several transitions from the $^3\Pi_{2g}$ ground state to the $^3\Sigma$ excited states are allowed and correspond to absorption bands near 880, 715, 470, and below 400 nm. Moreover, it was shown that NIR luminescence bands with wavelengths 1420–1520, 1275–1375, and 1000–1060 nm should also occur due to spin-forbidden transitions from the three $^1\Sigma$ singlet excited states to the $^3\Pi$ ground state and/or to one of the first two excited $^3\Pi$ states.[8] These calculated emission characteristics are similar to the experimentally determined PL spectrum (Figure 3). Despite this similarity, however, it is clear that at present the theoretical results are only approximate and do not completely agree with the experimental facts.

A better, although only qualitative, explanation of the observed photophysical behavior can be given as follows. The [Bi$_2$]$^{2-}$ anion absorbs photons with energies in the NIR and visible ranges as shown in Figure 5. After irradiation with high-energy photons, the electrons in the upper excited levels tend to nonradiatively relax to the first three excited levels from where the electrons relax to the lowest vibrational sublevels, thus resulting in Stokes NIR emissions. That is, the NIR PL is attributable to the radiative electronic transitions from the first three excited levels to the ground level based on a one-photon process. The overlapping of the three emission bands leads to the observed ultra-broad PL band ranging from 975 to 1400 nm.

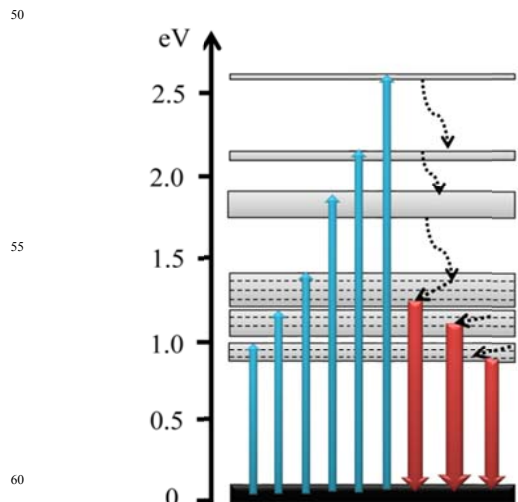

**Figure 5.** Simplified energy-level diagram of the [Bi$_2$]$^{2-}$ emitter based on the observed absorption spectrum from Figure 2. The blue and red arrows represent the excitation and emission bands, respectively. The black dotted arrows label the nonradiative relaxations. The dotted horizontal lines represent the vibrational sublevels corresponding to the lowest three excited levels.

In summary, we have demonstrated here that crystalline (K-crypt)$_2$Bi$_2$ emits ultra-broad NIR band, stemming from the inherent electronic transitions of [Bi$_2$]$^{2-}$. This represents the first experimental observation of NIR emission from negatively-charged bismuth species. These results not only add to the number of charged bismuth-based species with luminescence, but also deepen the understanding of bismuth related NIR emission behavior[4,5,8-14] as well as encourage efforts for the design of photonic materials with such emitters.

H. Sun gratefully acknowledges the funding support from Hokkaido University and NIMS, Japan. T. Yonezawa is grateful for the partial financial support through a Grant-in-Aid for Scientific Research (B) (21310072) from JSPS and a Grant-in-Aid for Scientific Research in Priority Area (Strong Photon-Molecule Coupling Fields for Chemical Reactions (470, 21020010) from MEXT, Japan. S. C. Sevov thanks the US National Science Foundation (CHE-0742365) for the financial support of this research. H. Sun greatly thanks the support from Dr. Z. H. Bai in Kobe University for the PL measurement.